\documentclass[12pt,prd,aps,amssymb,amsmath,tightenlines,showpacs]{revtex4}
\usepackage{psfig}

\usepackage{graphics}
\usepackage[pdftex]{graphicx}

\begin{document}
\preprint{}

\title{New Quasi-Exactly Solvable Sextic Polynomial Potentials}

\author{Carl~M.~Bender\footnote{Permanent address: Department of Physics,
Washington University, St. Louis, MO 63130, USA.} and Maria Monou}

\affiliation{Blackett Laboratory, Imperial College, London SW7 2BZ, UK}

\date{\today}

\begin{abstract}
A Hamiltonian is said to be {\it quasi-exactly solvable} (QES) if some of the
energy levels and the corresponding eigenfunctions can be calculated exactly and
in closed form. An entirely new class of QES Hamiltonians having sextic
polynomial potentials is constructed. These new Hamiltonians are different from
the sextic QES Hamiltonians in the literature because their eigenfunctions obey
$\mathcal{PT}$-symmetric rather than Hermitian boundary conditions. These new
Hamiltonians present a novel problem that is not encountered when the
Hamiltonian is Hermitian: It is necessary to distinguish between the parametric
region of unbroken $\mathcal{PT}$ symmetry, in which all of the eigenvalues are
real, and the region of broken $\mathcal{PT}$ symmetry, in which some of the
eigenvalues are complex. The precise location of the boundary between these two
regions is determined numerically using extrapolation techniques and
analytically using WKB analysis.
\end{abstract}


\maketitle

\section{SEXTIC QES HAMILTONIANS}
\label{sec1}

The purpose of this paper is to introduce a new class of quasi-exactly solvable
(QES) Hamiltonians having sextic polynomial potentials. While these new kinds of
QES Hamiltonians have positive, real eigenvalues, they have not yet been
discussed in the literature because they are not Hermitian. Instead, they are 
$\mathcal{PT}$ symmetric.

The term {\em quasi-exactly solvable} (QES) is used to describe a
quantum-mechanical Hamiltonian when a finite portion of its energy spectrum and
associated eigenfunctions can be found exactly and in closed form \cite{Ush}. 
Typically, QES potentials depend on a parameter $J$, and for positive integer
values of $J$ one can find exactly the first $J$ eigenvalues and eigenfunctions,
usually of a given parity. It has been shown that QES systems can be classified
by using an algebraic approach in which the Hamiltonian is expressed in terms of
the generators of a Lie algebra \cite{Tur,Tur1,ST,GKO}.

Perhaps the simplest example of a QES Hamiltonian having a sextic potential is
\cite{BD1,BD2}
\begin{equation}
H=p^2+x^6-(4J-1)x^2,
\label{eq1}
\end{equation}
where $J$ is a positive integer. For each positive integer value of $J$, the time-independent Schr\"odinger equation for this Hamiltonian,
\begin{equation}
-\psi''(x)+[x^6-(4J-1)x^2]\psi(x)=E\psi(x),
\label{eq2}
\end{equation}
has $J$ even-parity eigenfunctions in the form of an exponential times a
polynomial:
\begin{equation}
\psi(x)=e^{-x^4/4}\sum_{k=0}^{J-1}c_k x^{2k}.
\label{eq3}
\end{equation}
The polynomial coefficients $c_k$ ($0\leq k\leq J-1$) satisfy the recursion
relation
\begin{equation}
4(J-k)c_{k-1}+Ec_k+2(k+1)(2k+1)c_{k+1}=0,
\label{eq4}
\end{equation}
where we define $c_{-1}=c_{J}=0$. The simultaneous linear equations (\ref{eq4})
have a nontrivial solution for $c_0,\,c_1,\,...,\,c_{J-1}$ if the determinant of
the coefficients vanishes. This determinant is a polynomial of degree $J$ in the
variable $E$. The roots of this polynomial are all real and are the $J$
quasi-exact energy eigenvalues of the Hamiltonian (\ref{eq1}). Note that all of
the QES eigenfunctions (\ref{eq3}) of $H$ in (\ref{eq1}) have the form of a
decaying exponential $\exp(-\frac{1}{4}x^4)$ multiplying a polynomial. This is
the standard form in the literature for the eigenfunctions of any QES
Hamiltonian whose potential is a polynomial.

The QES Hamiltonians associated with Hermitian Hamiltonians have been examined
in depth and classified exhaustively \cite{Ush}. However, in 1998 new kinds of 
Hamiltonians that have positive real energy levels were discovered \cite{A,B}.
These new kinds of Hamiltonians are not Hermitian ($H\neq H^\dag$) in the usual
Dirac sense, where the Dirac adjoint symbol $\dag$ represents combined transpose
and complex conjugation. Instead, these Hamiltonians possess $\mathcal{PT}$
symmetry $H=H^{\mathcal{PT}}$; that is, they remain invariant under combined
space and time reflection. This new class of non-Hermitian Hamiltonians has been
studied heavily \cite{C,D,E} and it has been shown that when the $\mathcal{PT}$
symmetry is not broken, such Hamiltonians define unitary theories of quantum
mechanics \cite{BBJ}

The key difference between Hermitian Hamiltonians and complex, non-Hermitian,
$\mathcal{PT}$-symmetric Hamiltonians is that with $\mathcal{PT}$-symmetric
Hamiltonians the boundary conditions on the eigenfunctions (the solutions to the
time-independent Schr\"odinger equation) are imposed in wedges in the complex
plane. Sometimes these wedges do not include the real axis. (A detailed
discussion of the complex asymptotic behavior of solutions to eigenvalue
problems may be found in Ref.~\cite{BT}.)

The discovery of $\mathcal{PT}$-symmetric Hamiltonians was followed immediately
by the discovery of a new class of QES models. Until 1998 it was thought that
if the potential associated with a QES Hamiltonian was a polynomial, then this
polynomial had to be at least sextic; its degree could not be less than six.
This property is in fact true for Hamiltonians that are Hermitian. However, in
1998 it was discovered that it is possible to have a QES {\it non-Hermitian}
complex Hamiltonian whose potential is {\it quartic} \cite{BeBo}:
\begin{equation}
H=p^2-x^4+2iax^3+(a^2-2b)x^2+2i(ab-J)x.
\label{eq5}
\end{equation}
Here, $a$ and $b$ are real parameters and $J$ is a positive integer. For a large
region of the parameters $a$ and $b$, the energy levels of this family of
quartic Hamiltonians are real, discrete, and bounded below, and the quasi-exact
portion of the spectra consists of the lowest $J$ eigenvalues. Like the
eigenvalues of the Hamiltonian (\ref{eq1}), the lowest $J$ eigenvalues of these
potentials are the roots of a $J$th-degree polynomial \cite{KM}.

The reality of the eigenvalues of $H$ in (\ref{eq5}) is ensured by the boundary
conditions that its eigenfunctions are required to satisfy. The eigenfunctions
are required to vanish as $|x|\to\infty$ in the complex-$x$ plane inside of two
wedges called {\it Stokes wedges}. The right wedge is bounded above and below by
lines at $0^\circ$ and $-60^\circ$ and the left wedge is bounded above and below
by lines at $-180^\circ$ and $-120^\circ$. The leading asymptotic behavior of
the wave function inside these wedges is given by
\begin{eqnarray}
\psi(x)\sim e^{-ix^3/3}\quad(|x|\to\infty).
\label{eq6}
\end{eqnarray}

The new class of QES sextic Hamiltonians reported in this paper has the form
\begin{equation}
H=p^2+x^6+2ax^4+(4J-1+a^2)x^2,
\label{eq7}
\end{equation}
where $J$ is a positive integer and $a$ is a real parameter. These Hamiltonians
are very similar in structure to those in (\ref{eq1}) and to the other QES
sextic Hamiltonians discussed in the literature \cite{Ush}, but their
distinguishing characteristic is that the asymptotic behavior of their
eigenfunctions in the complex-$x$ plane is different.

Let us examine first the asymptotic behavior of the eigenfunction solutions to
the Schr\"odinger equation (\ref{eq2}). For brevity, we call the eigenfunctions
in (\ref{eq3}) the {\it good} solutions to (\ref{eq2}) because they satisfy the
physical requirement of being quadratically integrable. These good solutions
decay exponentially like $\exp(-\frac{1}{4}x^4)$ as $x\to\pm\infty$, while the
corresponding linearly independent {\it bad} solutions grow exponentially like
$\exp(\frac{1}{4}x^4)$ as $x\to\pm\infty$. In the complex-$x$ plane the good
solutions (\ref{eq3}) decay exponentially as $|x|\to\infty$ in two Stokes wedges
that are centered about the positive and the negative real-$x$ axes. These
wedges have an angular opening of $45^\circ$. The bad solutions grow
exponentially in these wedges. At the upper and lower edges of these wedges the
good and bad solutions cease to decay and to grow exponentially and they become
purely oscillatory.

As we move downward past the lower edges of these wedges, we enter a new pair of
Stokes wedges. These wedges also have a $45^\circ$ angular opening and are
centered about the lines ${\rm arg}\,x=-45^\circ$ and ${\rm arg}\,x=-135^\circ$.
In these lower wedges, the good solutions grow exponentially as $|x|\to\infty$
and thus they behave like a bad solutions.

In the lower pair of wedges we can find solutions to the new class of
Hamiltonians in (\ref{eq7}) that behave like good solutions. These new
$\mathcal{PT}$-symmetric eigenfunctions have the general form of the exponential
$\exp(\frac{1}{4}x^4+\frac{1}{2}ax^2)$ multiplied by a polynomial \cite{PARITY}:
\begin{equation}
\psi(x)=e^{x^4/4+ax^2/2}\sum_{k=0}^{J-1}c_k x^{2k}.
\label{eq8}
\end{equation}

Hamiltonians having even sextic polynomial potentials are special because such
Hamiltonians can be {\it either} Hermitian or $\mathcal{PT}$-symmetric depending
on whether the eigenfunctions are required to vanish exponentially in the
$45^\circ$ wedges containing the positive and negative real-$x$ axes or in the
other pair of $45^\circ$ wedges contiguous to and lying just below these wedges
in the complex-$x$ plane. The solutions for these two different boundary
conditions are somewhat related. Specifically, a good solution in one pair of
wedges becomes a bad solution in the other pair of wedges. However, a bad
solution in one pair of wedges does not become a good solution in the other
pair of wedges, as we now explain.

Given a good solution $\psi_{\rm good}(x)$ in one pair of wedges, we use the
method of reduction of order \cite{BO} to find the bad solution. We seek a bad
solution in the form $\psi_{\rm bad}(x)=\psi_{\rm good}(x)u(x)$, where $u(x)$ is
an unknown function to be determined. Substituting the bad solution into the
Schr\"odinger equation $-\psi''(x)+V(x)\psi(x)=E\psi(x)$, we get the
differential equation satisfied by $u(x)$:
\begin{equation}
\psi_{\rm good}(x)u''(x)+2\psi_{\rm good}'(x)u'(x)=0.
\label{eq9}
\end{equation}
We solve this equation by multiplying by the integrating factor $\psi_{\rm
good}(x)$ and obtain the result
\begin{equation}
\psi_{\rm bad}(x)=\psi_{\rm good}(x)\left(\int^x ds\,\left[\psi_{\rm
good}(s)\right]^{-2}+C\right),
\label{eq10}
\end{equation}
where $C$ is an arbitrary constant.

This bad solution always grows exponentially in the two wedges in which the good
solution decays exponentially. How does this bad solution behave in the other
pair of wedges in which the good solution grows exponentially? We can always
choose the constant $C$ so that the bad solution vanishes as $|x|\to\infty$ in
{\it one} of these two wedges. However, in the other of the two wedges, the bad
solution will always grow exponentially. Thus, while the good solution becomes
bad as we cross from one pair of wedges to the other, the bad solution does not
become good.

\section{Determination of the $\mathcal{PT}$ Boundary}
\label{sec2}

The difference between the Hermitian Hamiltonians in (\ref{eq1}) and the
$\mathcal{PT}$-symmetric Hamiltonians in (\ref{eq7}) is that the Hermitian
Hamiltonians always have real eigenvalues. The $\mathcal{PT}$-symmetric
Hamiltonians in (\ref{eq7}) have real eigenvalues only if the $\mathcal{PT}$
symmetry is unbroken; if the $\mathcal{PT}$ symmetry is broken, some of the
eigenvalues will be complex. Thus, it is crucial to determine whether the
$\mathcal{PT}$ symmetry is broken. We will see that there is a range of values
of the parameter $a$ in (\ref{eq7}) for which the energy levels are real and
this is the region of unbroken $\mathcal{PT}$ symmetry. Outside of this region
some of the eigenvalues appear as complex-conjugate pairs.

Let us illustrate the difference between the regions of broken and unbroken
$\mathcal{PT}$ symmetry by examining some special solutions of the Schr\"odinger
equation
\begin{equation}
-\psi''(x)+[x^6+2ax^4+(4J-1+a^2)x^2]\psi(x)=E\psi(x),
\label{eq11}
\end{equation}
corresponding to $H$ in (\ref{eq7}). First, consider the case $J=1$. The unique
eigenfunction solution to (\ref{eq11}) of the form in (\ref{eq8}) is $\psi(x)=
\exp(\frac{1}{4}x^4+\frac{1}{2}ax^2)$ and the corresponding eigenfunction is $E=
-a$. Note that $E$ is real so long as $a$ is real. Thus, for $J=1$ there is no
region of broken $\mathcal{PT}$ symmetry.

Next, consider the case $J=2$. Now, there are two eigenfunctions. The two
eigenvalues are given by
\begin{equation}
E=-3a\pm2\sqrt{a^2-2}.
\label{eq12}
\end{equation}
Thus, there is now an obvious transition between real eigenvalues (unbroken
$\mathcal{PT}$ symmetry) and complex eigenvalues (broken $\mathcal{PT}$
symmetry). Evidently, the eigenvalues are real if $a\geq \sqrt{2}$ or if $a\leq-
\sqrt{2}$. 

We find that for any positive integer value of $J>1$, the eigenvalues $E$ for
$H$ in (\ref{eq7}) are entirely real if $a^2$ is greater than some critical
value $[a_{\rm crit}(J)]^2$ that depends on $J$. These critical values up to $J=
20$ are shown in Table \ref{t1}.

\begin{table}[!hbtp]
\begin{center}
\begin{tabular}{c|c|c}
$J$\quad&\quad$[a_{\rm crit}(J)]^2$\quad&\quad$[a_{\rm crit}(J+1)]^2-[a_{\rm crit}(J)]^2$\quad\\
\hline
2   &  2 \\
3   &  10.5874700363 & \raisebox{1ex}{8.5874700363}\\
4   &  20.5515334397 & \raisebox{1ex}{9.9640634033}\\
5   &  31.0534552654 & \raisebox{1ex}{10.5019218257}\\
6   &  41.8519569727 & \raisebox{1ex}{10.7985017073}\\
7   &  52.8409390328 & \raisebox{1ex}{10.9889820601}\\
8   &  63.9636348939 & \raisebox{1ex}{11.1226958611}\\
9   &  75.1858755649 & \raisebox{1ex}{11.2222406710}\\
10  &  86.4853951835 & \raisebox{1ex}{11.2995196186}\\
11  &  97.8468072286 & \raisebox{1ex}{11.3614120451}\\
12  & 109.2590335351 & \raisebox{1ex}{11.4122263065}\\
13  & 120.7137913596 & \raisebox{1ex}{11.4547578245}\\
14  & 132.2047259144 & \raisebox{1ex}{11.4909345548}\\
15  & 143.7268461067 & \raisebox{1ex}{11.5221201923}\\
16  & 155.2761720922 & \raisebox{1ex}{11.5493064512}\\
17  & 166.8494020446 & \raisebox{1ex}{11.5732299524}\\
18  & 178.4439117241 & \raisebox{1ex}{11.5945096795}\\
19  & 190.0574079492 & \raisebox{1ex}{11.6134962251}\\
20\quad  &$\quad 201.6880273595\quad$ & \raisebox{1ex}{11.6306193103}\\
\end{tabular}  
\caption{Critical values, $[a_{\rm crit}(J)]^2$, of the parameter $a^2$ listed
as a function of $J$. When $a^2$ is greater than this critical value, the
eigenvalues of the $\mathcal{PT}$-symmetric Hamiltonian $H$ in (\ref{eq7}) are
all real. Thus, this is the region of unbroken $\mathcal{PT}$ symmetry. The
$\mathcal{PT}$ symmetry is broken when $a^2<[a_{\rm crit}(J)]^2$. Note that the
differences between successive values of $[a_{\rm crit}(J)]^2$ appear to be
approaching a limit and this is indeed the case. In fact, the numerical value of
this limit is exactly 12. Thus, for large $J$ the critical values have the
simple asymptotic behavior $[a_{\rm crit}(J)]^2\sim12J$.
\label{t1}}
\end{center}
\end{table}

\begin{figure}[t]
\vspace{3.20in}
\includegraphics{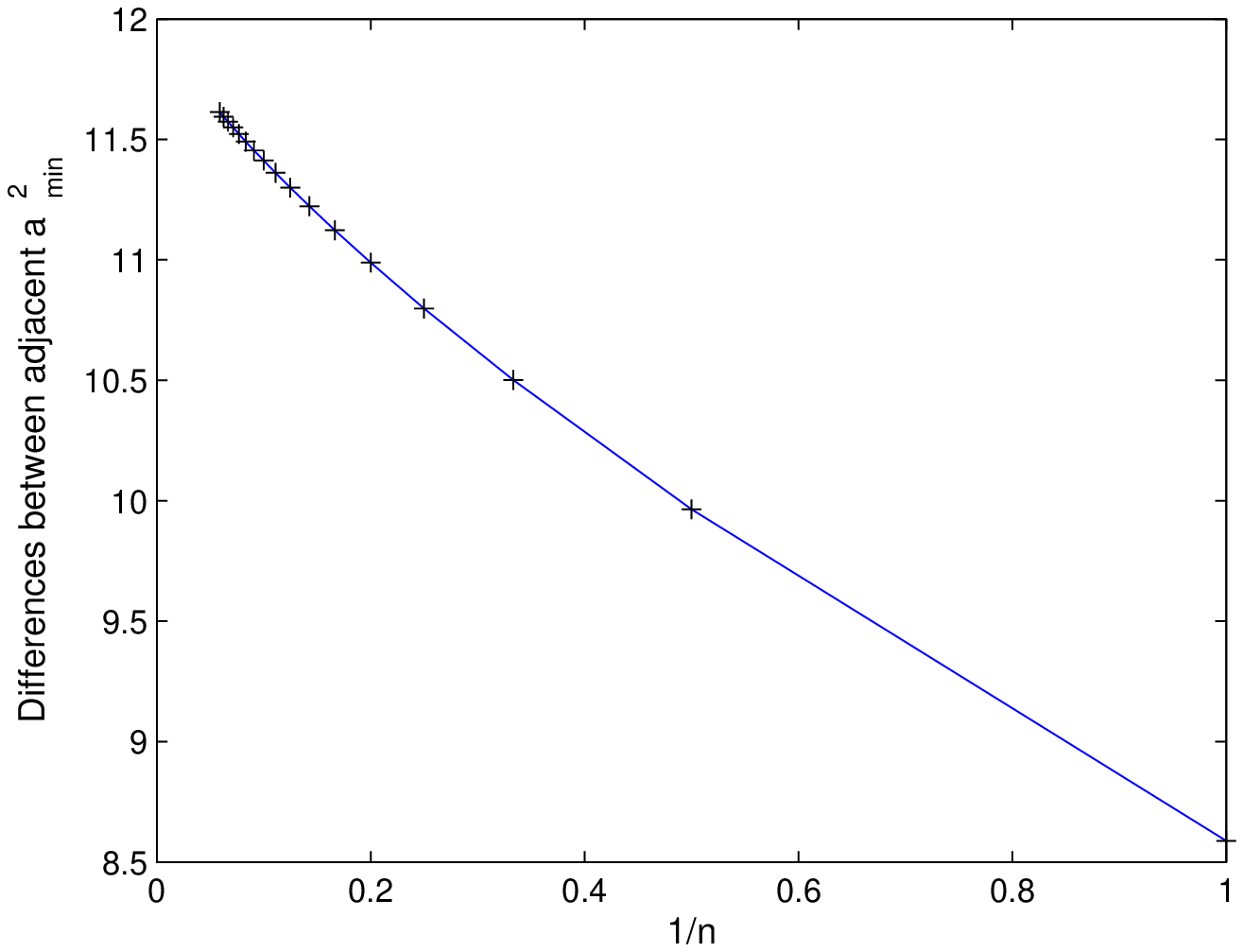}
\vspace{-17mm}
\caption{The differences $[a_{\rm crit}(J+1)]^2-[a_{\rm crit}(J)]^2$ taken from
Table \ref{t1} plotted as a function of $1/J$. Observe that as $J$ increases,
these differences tend to towards the limiting value $12$.}
\label{f1}
\end{figure}

Observe from Table \ref{t1} that the critical values of $[a_{\rm crit}]^2$ grow
monotonically with increasing $J$. We have therefore also calculated the
differences between successive critical values of $a^2$. These differences also
grow monotonically with increasing $J$, but they appear to be leveling off and
seem to be approaching a limiting value. To see whether the differences are
indeed approaching a limiting value as $J$ increases, we have plotted in
Fig.~\ref{f1} these differences as a function of $1/J$. This plot suggests that
the differences tend to the value $12$ as $J\to\infty$.

To determine whether it is true that these differences really do approach limit
$12$, it is necessary to extrapolate the sequence of differences to its value at
$J=\infty$. To do so we have calculated the Richardson extrapolants \cite{BO} of
the sequence of differences. The {\it first} Richardson extrapolants, $R_1(J)$,
of these differences are listed in Table \ref{t2}. Observe that the sequence
$R_1(J)$ rises more slowly and quite convincingly appears to be approaching the
value $12$. The differences $R_1(J+1)-R_1(J)$ between successive Richardson
extrapolants are also shown.

\begin{table}[!hbtp]
\begin{center}
\begin{tabular}{c|c|c}
$J$ &  $R_1(J)$ series & $R_1(J+1)-R_1(J)$ \\
\hline
1 & 11.3406567704\\
2 &11.5776386705  & \raisebox{1ex}{0.23698190}\\
3 &11.6882413518  & \raisebox{1ex}{0.11060268}\\ 
4 &11.7509034718  & \raisebox{1ex}{0.06266212}\\
5 &11.7912648657  & \raisebox{1ex}{0.04036140}\\
6 &11.8195095305  & \raisebox{1ex}{0.02824466}\\
7 &11.8404722516  & \raisebox{1ex}{0.02096272}\\
8 &11.8565514577  & \raisebox{1ex}{0.01607921}\\
9 &11.8695546582  & \raisebox{1ex}{0.01300320}\\
10 &11.8800730055 & \raisebox{1ex}{0.01051835}\\
11 &11.8888785336 & \raisebox{1ex}{0.00880552}\\
12 &11.8963479526 & \raisebox{1ex}{0.00746942}\\
13 &11.9027717144 & \raisebox{1ex}{0.00642376}\\
14 &11.9083609386 & \raisebox{1ex}{0.00558923}\\
15 &11.9132728866 & \raisebox{1ex}{0.00491195}\\
16 &11.9176271918 & \raisebox{1ex}{0.00435430}\\
\end{tabular}
\caption{First Richardson extrapolants $R_1(J)$ of the sequence of differences
$[a_{\rm crit}(J+1)]^2-[a_{\rm crit}(J)]^2$ taken from Table \ref{t1}. Notice
that $R_1(J)$ rises slowly and smoothly towards its limiting value $12$. The
differences between successive Richardson extrapolants are also listed.
\label{t2}}
\end{center}
\end{table}

To test further the hypothesis that $R_1(J)$ tends to the limiting value $12$ as
$J\to\infty$, we have calculated successive Richardson extrapolants of the
Richardson extrapolants $R_1(J)$ in Table \ref{t2}. The successive extrapolants
are listed in Table \ref{t3} and they provide very strong numerical evidence
that $\lim_{J\to\infty}\left([a_{\rm crit}(J+1)]^2-[a_{\rm crit}(J)]^2\right)=
12$. From this we conclude that for large $J$ the asymptotic behavior of the
critical value of $a^2$ is given by
\begin{equation}
[a_{\rm crit}(J)]^2\sim12J\quad(J\to\infty).
\label{eq13}
\end{equation}

\begin{table}[!hbt]
\begin{center}
\begin{tabular}{c|c|c|c|c} 
J &  $R_1(J)$ & R of $R_1(J)$ & R of R of $R_1(J)$ & R of R of R of $R_1(J)$ \\
\hline
1 & 11.3406567704 & 11.8146205706 & 12.0042728584 & 11.9912792745 \\ 
2 & 11.5776386705 & 11.9094467145 & 11.9977760665 & 11.9869646719 \\
3 & 11.6882413518 & 11.9388898318 & 11.9941722683 & 11.9887732331 \\
4 & 11.7509034718 & 11.9527104409 & 11.9928225095 & 11.9978459499 \\
5 & 11.7912648657 & 11.9607328547 & 11.9938271976 & 11.9483630075 \\
6 & 11.8195095305 & 11.9662485785 & 11.9862498326 & 12.1168065542 \\
7 & 11.8404722516 & 11.9691059005 & 12.0049007928 & 11.8377031045 \\
8 & 11.8565514577 & 11.9735802620 & 11.9840010818 & 12.0982460411 \\
9 & 11.8695546582 & 11.9747381309 & 11.9966949661 & 11.9726356846 \\
10& 11.8800730055 & 11.9769338144 & 11.9942890380 & 11.9983156753 \\
11& 11.8888785336 & 11.9785115620 & 11.9946550959 \\
12& 11.8963479526 & 11.9798568565 \\
13& 11.9027717144 \\
\end{tabular}
\caption{Repeated Richardson extrapolants of the sequence of Richardson
extrapolants in Table \ref{t2}. This table provides strong and convincing
numerical evidence that Richardson extrapolants $R_1(J)$ tend to the limiting
value $12$ as $J\to\infty$. This implies that for large $J$ the critical values
of $a^2$ grow linearly with $J$. See Eq.~(\ref{eq13}).
\label{t3}}
\end{center} 
\end{table}

Our numerical analysis provides convincing evidence that for large $J$ the
boundary between the regions of broken and unbroken $\mathcal{PT}$ symmetry is
given by the asymptotic behavior in (\ref{eq13}). We will now verify this
result analytically by using WKB methods \cite{BO}. From our numerical analysis
we know that the first eigenvalues to become complex conjugate pairs are always
the highest, and this implies that WKB is the appropriate tool for investigating
the $\mathcal{PT}$ boundary for large $J$.

For the potential $V(x)=x^6+2ax^4+(a^2+4J-1)x^2$, the leading-order WKB
quantization condition, valid for large $n$, is
\begin{equation}
(2n+\textstyle{\frac{1}{2}})\pi\sim\int_{T_1}^{T_2}dx\sqrt{E_n-V(x)}\quad
(n\to\infty),
\label{eq14}
\end{equation}
where $T_{1,2}$ are the turning points. Note that there is a factor of $2n+
\frac{1}{2}$, rather than $n+\frac{1}{2}$, on the left side of this asymptotic
relation because we are counting {\it even}-parity eigenfunctions.

For large $n=J$ we approximate the integral in (\ref{eq14}) by making the
asymptotic substitution $a\sim\sqrt{J}b$, where $b$ is a number to be
determined. In order to verify the asymptotic behavior in (\ref{eq13}), we
must show that $b=\sqrt{12}$. We then make the scaling substitutions
\begin{equation}
x=yJ^{1/4}\quad{\rm and}\quad E_J\sim FJ^{3/2}
\label{eq15}
\end{equation}
because for large $J$ we can then completely eliminate all dependence on $J$
from the integral. We thus obtain the condition
\begin{equation}
2\pi=\int_{y=U_1}^{U_2}dy\sqrt{F-[y^6+2by^4+(b^2+4)y^2]},
\label{eq16}
\end{equation}
where $U_{1,2}=T_{1,2}J^{-1/4}$ are zeros of the algebraic equation
\begin{equation}
y^6+2by^4+(b^2+4)y^2-F=0.
\label{eq17}
\end{equation}

Next, following the analysis in Ref.~\cite{BD2}, we assume that in this
large-$J$ limit the polynomial in (\ref{eq17}) factors:
\begin{equation}
(y^2-\alpha)^2(y^2-\beta)=0.
\label{eq18}
\end{equation}
The correctness of this factorization assumption will be verified in the
subsequent analysis. We then expand (\ref{eq18});
\begin{equation}
y^6-y^4(\beta+2\alpha)+y^2(\alpha^2+2\alpha\beta)-\alpha^2\beta=0.
\label{eq19}
\end{equation}
Comparing coefficients of like powers of $y$ in (\ref{eq17}) and (\ref{eq19}),
we obtain the three equations
\begin{equation}
F=\alpha^2\beta,
\label{eq20}
\end{equation}
\begin{equation}
2b=-2\alpha-\beta,
\label{eq21}
\end{equation}
\begin{equation}
b^2+4=\alpha^2+2\alpha\beta.
\label{eq22}
\end{equation}

Subtracting the square of Eq. (\ref{eq21}) from three times (\ref{eq22}), we
get $\beta-\alpha=\pm\sqrt{b^2-12}$, and solving this equation simultaneously
with (\ref{eq21}), we get expressions for $\alpha$ and $\beta$:
\begin{equation}
3\alpha=-2b-\sqrt{b^2-12},
\label{eq23}
\end{equation}
\begin{equation}
3\beta=-2b+2\sqrt{b^2-12}.
\label{eq24}
\end{equation}
We then substitute (\ref{eq23}) and (\ref{eq24}) into (\ref{eq20}) to obtain
\begin{equation}
F=-\textstyle{\frac{2}{27}}(b-\sqrt{b^2-12})(2b+\sqrt{b^2-12})^2.
\label{eq25}
\end{equation}

Finally, we calculate the value of the number $b$. Our procedure is simply to
show that the special choice $b^2=12$ is consistent with the limiting WKB
integral in (\ref{eq16}). With this choice we can see from (\ref{eq23}) and
(\ref{eq24}) that $\alpha=\beta=4/\sqrt{3}$ and that (\ref{eq16}) reduces to
\begin{equation}
2\pi=\int_{y=-\alpha}^{\alpha}dy\,(\alpha-y^2)^{3/2}=
2\int_{y=0}^{\alpha}dy\,(\alpha-y^2)^{3/2}.
\label{eq26}
\end{equation}
We simplify this integral by making the substitution $y=\sqrt{u\alpha}$, and
obtain
\begin{equation}
\textstyle{\frac{3}{8}}\pi=\int_{u=0}^1 du\,u^{-1/2}(1-u)^{3/2},
\label{eq27}
\end{equation}
which is an exact identity. Thus, we may conclude that $b^2=12$. This verifies
the asymptotic formula in (\ref{eq13}) for the location of the $\mathcal{PT}$
boundary.

Furthermore, we can see that $F=\frac{64}{9}\sqrt{3}\approx12.3$. Thus, we
obtain a formula for the large-$J$ asymptotic behavior of the largest QES
eigenvalue at the $\mathcal{PT}$ boundary:
\begin{equation}
E_J\sim\textstyle{\frac{64}{9}}\sqrt{3}J^{3/2}\quad(J\to\infty).
\label{eq28}
\end{equation}

The difference between this WKB calculation and that done in Ref.~\cite{BD2}
for the Hermitian QES sextic Hamiltonian (\ref{eq1}) is that here we have a
critical value, $b=\sqrt{12}$, or $a\sim\sqrt{12J}$. This critical value defines
the boundary between the regions of broken and unbroken $\mathcal{PT}$ symmetry
for the $\mathcal{PT}$-symmetric Hamiltonian in (\ref{eq7}). There is no
analog of this boundary for Hermitian Hamiltonians.

\begin{acknowledgments}
We are greatful to Dr. H. F. Jones for giving us valuable advice with regard to
our WKB approximations. CMB is grateful to the Theoretical Physics Group at
Imperial College for its hospitality and he thanks the U.K. Engineering and
Physical Sciences Research Council, the John Simon Guggenheim Foundation, and
the U.S.~Department of Energy for financial support. MM gratefully acknowledges
the financial support of ???.
\end{acknowledgments}


\begin{enumerate}

\bibitem{Ush} See A.~G.~Ushveridze, {\sl Quasi-Exactly Solvable Models in
Quantum Mechanics} (Institute of Physics, Bristol, 1993) and references therein.

\bibitem{Tur} A.~V.~Turbiner, Sov.~Phys., JETP {\bf 67}, 230 (1988),
Contemp.~Math.~{\bf 160}, 263 (1994), and
M.~A.~Shifman, Contemp.~Math.~{\bf 160}, 237 (1994).

\bibitem{Tur1}
A.~V.~Turbiner, Comm. Math. Phys.~{\bf 118}, 467 (1988).

\bibitem{ST} M.~A.~Shifman and A.~V.~Turbiner, Comm. Math. Phys. {\bf 126},
347 (1989).

\bibitem{GKO} A.~Gonz\'alez-L\'opez, N.~Kamran, and P.~J.~Olver, Comm.
Math.~Phys.~{\bf 153}, 117 (1993) and Contemp.~Math.~{\bf 160}, 113 (1994).

\bibitem{BD1} C.~M.~Bender and G.~V.~Dunne, J.~Math.~Phys.~{\bf 37}, 6 (1996).

\bibitem{BD2} C.~M.~Bender, G.~V.~Dunne, and M.~Moshe, Phys.~Rev.~A {\bf 55},
2625 (1997).

\bibitem{A} C.~M.~Bender and S.~Boettcher, Phys. Rev.~Lett. {\bf 80}, 5243-5246
(1998).

\bibitem{B} C. M. Bender, S. Boettcher, and P. N. Meisinger,
J. Math. Phys. {\bf 40}, 2201-2229 (1999).

\bibitem{C} P.~Dorey, C.~Dunning and R.~Tateo, J.~Phys.~A {\bf 34} L391
(2001); {\em ibid}. {\bf 34}, 5679 (2001).

\bibitem{D} G.~L\'evai and M.~Znojil, J.~Phys. A{\bf 33}, 7165 (2000); B.~Bagchi
and C.~Quesne, Phys.~Lett. A{\bf 300}, 18 (2002); Z. Ahmed, Phys. Lett. A{\bf
294}, 287 (2002); G.~S.~Japaridze, J.~Phys.~A{\bf 35}, 1709 (2002);
A.~Mostafazadeh, J.~Math.~Phys. {\bf 43}, 205 (2002); {\em ibid}; {\bf 43}, 2814
(2002); D.~T.~Trinh, PhD Thesis, University of Nice-Sophia Antipolis (2002), and
references therein.

\bibitem{E} An excellent summary of the current status and the background of
non-Hermitian and $\mathcal{PT}$-symmetric Hamiltonians may be found in
F.~Kleefeld, hep-th/0408028 and hep-th/0408097.

\bibitem{BBJ} C. M. Bender, D. C. Brody, and H. F. Jones, Phys. Rev. Lett. {\bf
89}, 270401 (2002) and Am. J. Phys. {\bf 71}, 1095 (2003).

\bibitem{BT} C.~M.~Bender and A.~Turbiner, Phys.~Lett.~A {\bf 173}, 442 (1993).

\bibitem{BeBo} C. M. Bender and S. Boettcher, J. Phys. A: Math. Gen. {\bf 31},
L273 (1998).

\bibitem{KM} For a nonpolynomial QES $\mathcal{PT}$-symmetric Hamiltonian see
A.~Khare and B.~P.~Mandal, Phys.~Lett.~A {\bf 272}, 53 (2000).

\bibitem{PARITY} Notice that $\psi(x)$ in (\ref{eq8}) is a function of $x^2$ and
thus all of the QES wave functions are symmetric under parity reflection ($x\to-
x$). In general, $\mathcal{PT}$-symmetric Hamiltonians, such as $H=p^2-x^4$ are
not symmetric under parity reflection because the parity operator $\mathcal{P}$ 
changes the complex domain of the Hamiltonian operator. As a consequence, the
expectation value of the $x$ operator is nonvanishing. [See C.~M.~Bender,
P.~N.~Meisinger, and H.~Yang, Phys.~Rev.~D 63, 45001 (2001).] Nevertheless, the
special QES eigenfunctions in (\ref{eq8}) {\it are} parity-symmetric. We believe
that the parity operator may therefore be used to distinguish between the QES
and the non-QES portions of the Hilbert space.

\bibitem{BO} C.~M.~Bender and S.~A.~Orszag, {\it Advanced Mathematical Methods
for Scientists and Engineers}, (McGraw-Hill, New York, 1978), Chap.~10.

\end{enumerate}
\end{document}